\documentclass[5p, twocolumn]{elsarticle}
\usepackage{graphicx}
\usepackage{dcolumn}
\usepackage{txfonts}
\usepackage{float}
\usepackage{lineno}
\usepackage{bm}
\usepackage[ulem=normalem]{changes}
\usepackage{ulem}
\usepackage{todonotes} 
\usepackage{color}
\usepackage{booktabs}
\usepackage{stackrel}
\usepackage{upgreek}

\newcommand{\Fe}{$^{52}$Fe}
\newcommand{\Ni}{$^{56}$Ni}
\newcommand{\Co}{$^{54}$Co\,}
\newcommand{\Mn}{$^{50}$Mn\,}
\newcommand{\phe}{(p,$^{3}$He)}

\def \micro#1{\upmu#1}   

\begin{document}
\title{ Neutron-proton pairing in the N=Z radioactive \textit{fp}-shell nuclei $^{56}$Ni and $^{52}$Fe probed by pair transfer. }
\author[1]{B. Le Crom}
\author[1]{M. Assi\'e \corref{cor1}}
\ead{marlene.assie@ijclab.in2p3.fr}
\author[1]{Y. Blumenfeld}
\author[1]{ J. Guillot }
\author[2]{H. Sagawa}
\author[3]{T. Suzuki}
\author[2]{M. Honma}
\author[4]{N.L. Achouri}
\author[5]{ B. Bastin }
\author[6]{R. Borcea}
\author[7]{ W.N. Catford}
\author[5]{ E. Cl\'ement}
\author[5]{ L. C\'aceres}
\author[8]{ M. Caama\~no} 
\author[9]{ A. Corsi}
\author[5]{ G. De France }
\author[4]{ F. Delaunay }
\author[1]{N. De S\'er\'eville}
\author[8]{ B. Fernandez-Dominguez}
\author[10] {M. Fisichella}
\author[1]{ S. Franchoo }
\author[1]{A. Georgiadou}
\author[4]{J. Gibelin }
\author[9]{ A. Gillibert }
\author[1]{ F. Hammache }
\author[5]{ O. Kamalou }
\author[7]{  A. Knapton }
\author[9]{ V. Lapoux} 
\author[4]{ S. Leblond }
\author[11]{A.O. Macchiavelli}
\author[4]{ F.M. Marqu\'es }
\author[7]{ A. Matta \fnref{fn1}}
\author[5]{ L. M\'enager }
\author[1]{ P. Morfouace \fnref{fn2} }
\author[4]{ N.A. Orr }
\author[5]{ J. Pancin }
\author[4,8]{ X. Pereira-Lopez }
\author[1]{L. Perrot }
\author[5]{ J. Piot}
 \author[9]{  E. Pollacco }
 \author[8]{  D. Ramos  \fnref{fn3} }
\author[5]{T. Roger }
\author[6]{ F. Rotaru }
\author[12]{A. M. S\'anchez-Ben\'{\i}tez \fnref{fn4}}
\author[9]{ M. S\'enoville}
\author[5]{O. Sorlin }
\author[6]{ M. Stanoiu }
\author[1]{  I. Stefan }
\author[5]{ C. Stodel}
\author[1]{ D. Suzuki \fnref{fn5}}
\author[5]{ J-C Thomas}
\author[5]{ M. Vandebrouck \fnref{fn6}}
\cortext[cor1]{Corresponding author}
\fntext[fn1]{Present address: Laboratoire de Physique Corpusculaire de Caen, ENSICAEN  CNRS/IN2P3, 14050 Caen, France}
\fntext[fn2]{Present address: Commissariat \`a l'Energie Atomique, Service de Physique Nucl\'eaire, BP 12, F-91680 Bruy\`eres-le-Ch\^atel, France}
\fntext[fn3]{Present address: Grand Acc\'el\'erateur National d’Ions Lourds (GANIL), CEA/DRF-CNRS/IN2P3, Bvd Henri Becquerel, 14076 Caen, France}
\fntext[fn4]{Present address:Centro de Estudios Avanzados en F\'{\i}sica, Matem\'aticas y Computaci\'on (CEAFMC), Department of Integrated Sciences, University of Huelva, 21071 Huelva, Spain}
\fntext[fn5]{Present address: RIKEN Nishina Center, 2-1 Hirosawa, Wako, Saitama 351-0198, Japan}
\fntext[fn6]{Present address: IRFU, CEA, Universit\'e Paris-Saclay, F-91191 Gif-sur-Yvette, France }

\address[1]{Universit\'e Paris-Saclay, CNRS/IN2P3, IJCLab, 91405 Orsay, France }
\address[2]{Center for Mathematical Sciences, University of Aizu, Aizu-Wakamatsu, Fukushima 965-8580, Japan}
\address[3]{Department of Physics, College of Humanities and Sciences, Nihon University, Sakurajosui 3-25-40, Setagaya-ku, Tokyo 156-8550, Japan}
\address[4]{Laboratoire de Physique Corpusculaire de Caen, ENSICAEN  CNRS/IN2P3, 14050 Caen, France}
\address[5]{Grand Acc\'el\'erateur National d’Ions Lourds (GANIL), CEA/DRF-CNRS/IN2P3, Bvd Henri Becquerel, 14076 Caen, France}
\address[6]{Horia Hulubei National Institute of Physics and Nuclear Engineering, Magurele, Romania}
\address[7]{Department of Physics, University of Surrey, Guildford GU2 5XH, United Kingdom}
\address[8]{IGFAE and Dpt. de F\'{i}sica de Part\'{i}culas, Univ. of Santiago de Compostela, E-15758, Santiago de Compostela, Spain}
\address[9]{IRFU, CEA, Universit\'e Paris-Saclay, F-91191 Gif-sur-Yvette, France}
\address[10]{Laboratori Nazionali del Sud, Instituto Nazionale di Fisica Nucleare, Catania, Italy}
\address[11]{Nuclear Science Division, Lawrence Berkeley National Laboratory, Berkeley, CA 94720, USA}
\address[12]{Nuclear Physics Center, University of Lisbon, P-1649-003 Lisbon, Portugal}

%
\begin{abstract}
The isovector and isoscalar components of neutron-proton pairing are investigated in the N=Z unstable nuclei of the \textit{fp}-shell through the two-nucleon transfer reaction (p,$^3$He) in inverse kinematics. The combination of particle and gamma-ray detection with radioactive beams of $^{56}$Ni and $^{52}$Fe, produced by fragmentation at the GANIL/LISE facility, made it possible to carry out this study for the first time in a closed and an open-shell nucleus in the \textit{fp}-shell. 
The transfer cross-sections for ground-state to ground-state (J=0$^+$,T=1) and to the first (J=1$^+$,T=0) state were extracted for both cases  together with the transfer cross-section ratios $\sigma$(0$^+$,T=1) /$\sigma$(1$^+$,T=0). They are compared with second-order distorted-wave born approximation (DWBA) calculations.  
The enhancement of the ground-state to ground-state pair transfer cross-section close to mid-shell, in $^{52}$Fe, points towards a superfluid phase in the isovector channel.
For the "deuteron-like" transfer, very low cross-sections to the first (J=1$^+$,T=0) state were observed both for \Ni\phe\, and \Fe\phe\, and are related to a strong hindrance of this channel due to spin-orbit effect. No evidence for an isoscalar deuteron-like condensate is observed.

  \end{abstract}
%

\maketitle

Pairing correlations are at the origin of superfluidity in strongly interacting quantum many-body systems~\cite{Bri05,Bro14}. 
The theoretical description of these correlations is rooted in the microscopic theory of superconductivity developed by Bardeen, Cooper and Schrieffer (BCS) \cite{BCS} with building blocks made of strongly correlated pairs, the Cooper pairs. For most of the known nuclei, the superfluid states consist of isovector (T=1) neutron and/or proton pairs (nn or pp pairs). In N=Z nuclei, the large overlap between neutron and proton wave functions allows another type of Cooper pairs made of neutron-proton pairs (np pairs) of two different types: either isovector Cooper pairs (L=0, S=0, T=1) as for the nn and pp pairs or isoscalar Cooper pairs (L=0, S=1, T=0), the "deuteron-like" pairing. The latter is a very unique manifestation in nature of cross-species pairing \cite{Fra15} .

Cooper pair transfer is an efficient probe to unravel the properties of the pairing force. 
Systematic neutron pair transfer (adding or removing a pair), performed within a major shell, reveals the occurence of a paired superfluid phase. 
For open-shell nuclei, like stable singly-magic Sn isotopes, the pair transfer amplitude is enhanced at mid-shell and is driven by the ratio of the pairing gap to the strength of the pairing force, which reflects the number of Cooper pairs contributing coherently \cite{Bri05,Bro14,Yos62}. For closed-shell nuclei, like $^{208}$Pb, pair vibrations sign the onset of the superfluid phase and the pair transfer cross-section is governed by the number of phonons involved \cite{Bri05,Bro14,Mot75}. 
%

Similar arguments on the pair transfer cross-sections hold for np pair \cite{Fro71, Isa05, Alex18}.  In this case, two components enter into play: the isoscalar pair and the isovector pair transfer. Their interplay can be probed starting from an even-even self-conjugate nucleus (J=0$^+$,T=0) and populating respectively the lower lying (J=1$^+$,T=0) and (J=0$^+$,T=1) states in the residual odd-odd nucleus. The relative strength of the isoscalar and isovector channel is obtained from the ratio of the cross-sections: $\sigma$(0$^{+}$,T=1)/$\sigma$(1$^{+}$,T=0). 

The np pair stripping and pick-up reactions performed in direct kinematics in stable N=Z nuclei of the \textit{sd}-shell have been reviewed in \cite{Fra15} and recently remeasured consistently \cite{Ayy17}. Isoscalar pairing remains elusive in the \textit{sd}-shell due, perhaps, to the limited possible occupation and calls  for further studies in heavier nuclei. Indeed, high-j orbitals are more favorable as the typical number of pairs entering into play depends on the degeneracy of the specific orbitals involved. Refs.~\cite{Ber10, Alex11}  have shown that the region of N=Z$\approx$64 is the most promising to observe a competition between the two channels. This region is out of reach today for transfer reaction studies but higher \textit{j} orbitals and/or higher shells where N=Z nuclei are unstable can be tackled. These are very challenging measurements given the low cross-sections, the radioactive beam intensities available and the high density of states in the residual odd-odd nuclei. Indeed, only one attempt was made so far to measure $^{44}$Ti($^3$He,p)$^{46}$V with a radioactive (long-lived) $^{44}$Ti beam \cite{Fra15,aom}.
In the present work, we push the measurements up to the end of the \textit{f}$_{7/2}$ shell by studying the np pair transfer reactions \Ni\phe\Co and \Fe\phe\Mn, involving respectively a doubly magic nucleus and a near mid-shell nucleus to investigate the relative strength of the isovector and isoscalar Cooper pairs.

For these heavier nuclei, the LS coupling does not hold anymore due to the effect of the spin-orbit force. In the \textit{jj}-coupling, the isovector pairing concentrates in the J=0 channel whereas for the isoscalar pairing, two main couplings are favored: the anti-aligned (J=1) and the maximum aligned (J=J$_{max}$=7 in our case) \cite{Sch76}. Thus the cross-section to the (J=0$^+$,T=1) state reflects the contribution of np Cooper pairs in the isovector channel whereas the cross-section to the (J=1$^+$,T=0) state reflects the isoscalar np pairs with their two main components: the Cooper pairs (J=1) and the aligned pairs (J=7). To discuss the contribution of the latter, we assume a simplified picture with only the \textit{f}$_{7/2}$ orbit. For \Ni\phe, the transferred pair can only be a neutron and a proton holes coupled to J=1, so that there is no aligned np pairs contribution. In the \Fe\phe\,case, IBM calculations \cite{Isa13} show that the ground state of $^{52}$Fe can be described as two aligned np pairs coupled to J=0, but the lower lying (J=1$^+$,T=0) state of $^{50}$Mn has very little overlap with 3 aligned np pairs coupled to J=1. Therefore, the aligned np pairs contribution plays a minimal role in both reactions. Thus, the present study concentrates on isovector and isoscalar Cooper pairs.

%

%
%

The experiment was performed at GANIL \cite{GAN}. The beams of \Ni~and \Fe~were produced at 30.5A MeV and 31.2A MeV respectively, with an intensity of 10$^5$pps, by fragmentation of a $^{58}$Ni primary beam accelerated to 74.5A MeV on a 1\,mm-thick Be target. Two different settings of the LISE spectrometer \cite{Ann} including the Wien filter were used to select each beam.  Two low-pressure multi-wire devices, CATS \cite{cats}, were placed 
upstream of the target in order to reconstruct event by event the beam trajectory and its position on the target.  They also provide beam time reference and precise determination of the number of incoming ions. 
Beam contaminants, which represented 40\% of the total intensity, were removed offline using a time-of-flight measurement between the cyclotron radio-frequency and the first CATS detector.
The secondary beams impinged on a 6.8 mg/cm$^{2}$ thick CH$_{2}$ target, in which (p,$^3$He) reactions occurred. A $^{12}$C target was also used to estimate the carbon background. 
 \begin{figure}[h!]
\begin{center}
\includegraphics[height=.2\textheight]{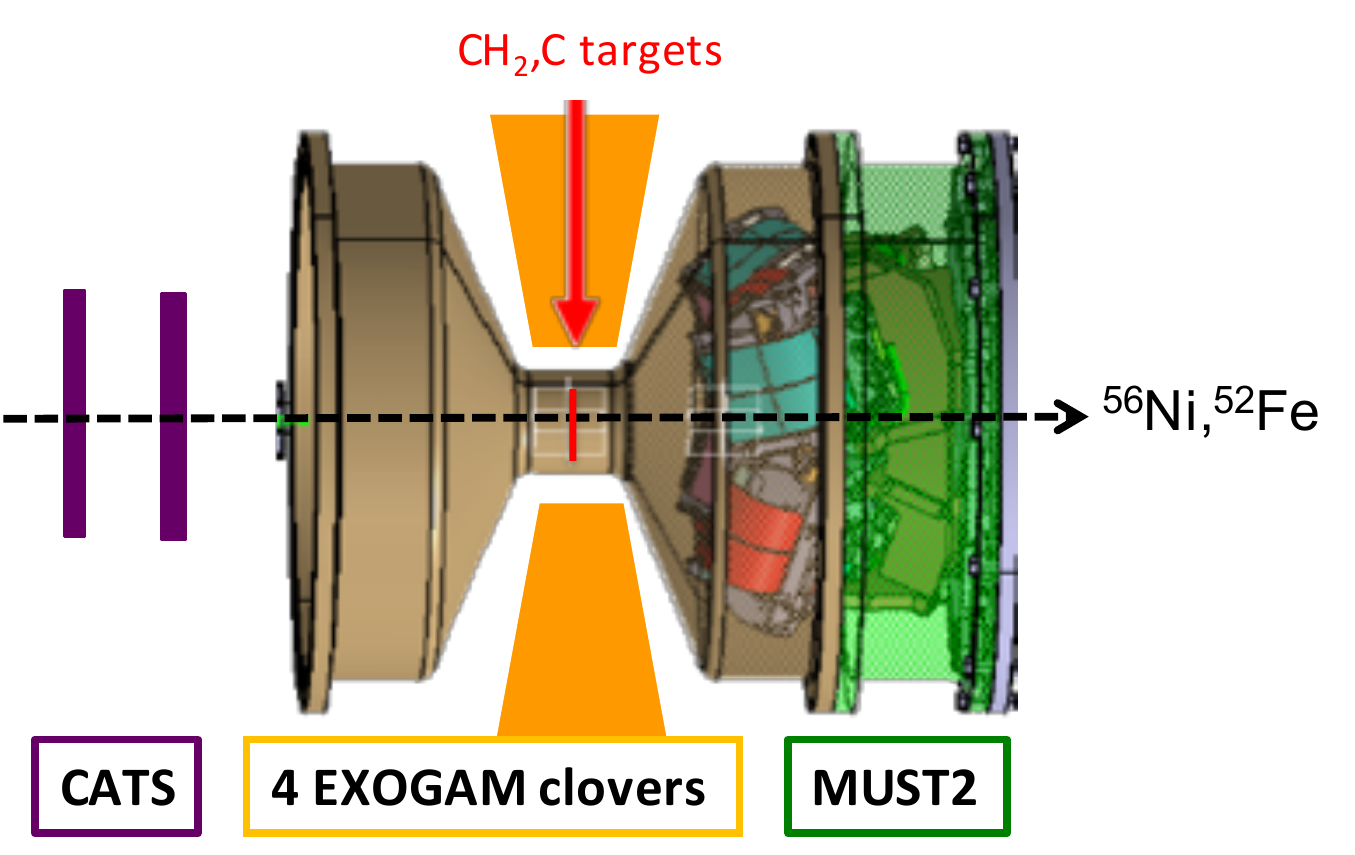}
\caption{ Experimental set-up (see text for details).\label{setup}}
\end{center}
\end{figure}

\begin{figure*}[h]
{\includegraphics[width=1.0\linewidth]{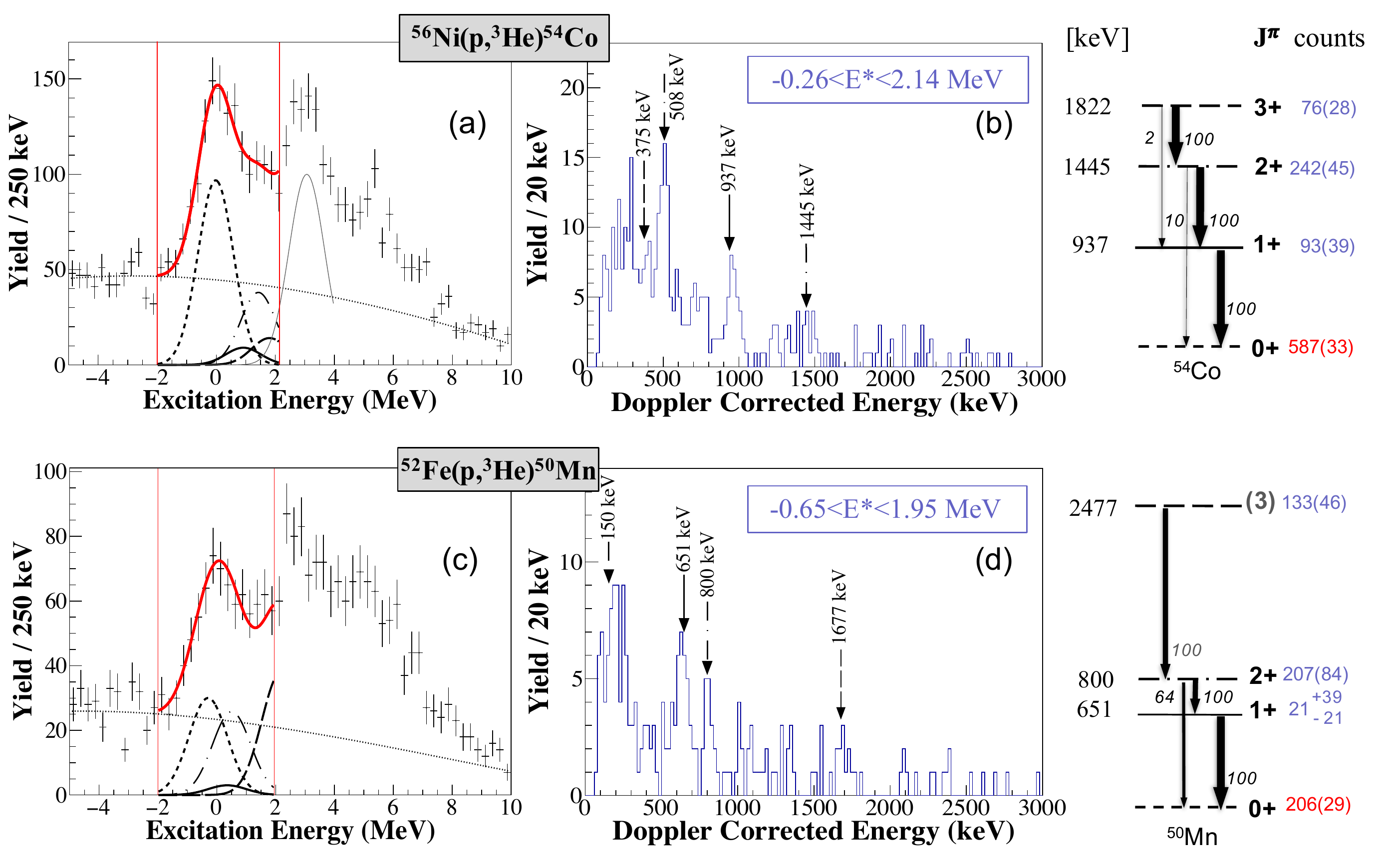}}
\caption{ (Color online) (Left) Excitation energy spectra for \Ni\phe\Co (a) and for \Fe\phe\Mn (c). The background contribution deduced from the measurement on a $^{12}$C target is shown as a light dotted line (see text for details). (Right) The associated gamma spectra with a gate on the excitation energy centered on the first (J=1$^+$,T=0) state $\pm$2$\sigma$ (the gate in E$^*$ is given explicitly on the spectra). A simplified level scheme of the residual nuclei is shown with the main transitions and associated intensities taken from \cite{ENSDF}. The deduced direct feeding by the transfer reaction (in number of counts) of each excited state is shown in blue with its statistical error bars (see text for details). This information is used to constrain the fit of the excitation energy (shown as a thick red line on spectra a) and c)). The population of the ground state (given in red) is deduced from the fit. The contribution of each state to the fit is identified with the same line code as for the associated gamma-ray lines (dotted for the gs, full for the 1$^+$, dashed dotted for the 2$^+$ and dashed for the 3$^{(+)}$ state). For \Ni\phe\Co a contribution at higher energy (around 3 MeV) is also shown and detailed in the text.\label{gamma}}
\end{figure*}

The energies and angles of the emitted light ejectiles were measured by four MUST2 telescopes \cite{must2} in the forward direction covering from 10 to 40 degrees in the laboratory frame (see Fig.\,\ref{setup}). 
Each telescope consists of a 300 $\mu$m-thick double-sided stripped Silicon detector (DSSD) with 128 strips on each side backed by 16 4 cm-thick CsI scintillators read out by photodiode which provide energy-loss ($\Delta$E) and residual energy (E) measurements, respectively.  The DSSD were calibrated with a triple alpha source and the CsI calibration for $^{3}$He and deuterons is deduced from their energy losses in the DSSD. The intrinsic energy resolution is of 40 keV (FWHM) for the former and  0.08$\sqrt{E_{CsI}}$ MeV for the latter. 
Light particle identification was obtained from the $\Delta$E-E measurement. 
The emission angle of the light ejectiles was deduced from the beam direction and impact position on the target given by the two CATS detectors and from the position information given by the DSSD with a precision better than 1 degree. Its energy was corrected for the energy losses in half of the target and in the dead-layers of the DSSD. 

The calibration and alignement of the detectors were validated with the one-neutron transfer reaction \Ni(p,d)$^{55}$Ni  where the ground state excitation energy is reconstructed at $-40\pm$50 keV with an energy resolution of $\sigma$=420 keV. This reaction was also used as a test bench for simulations of the experimental set-up performed with the \emph{nptool} package \cite{nptool}. With realistic conditions for beam spread and beam reconstruction, an energy resolution of $\sigma$=390 keV was found, confirming the reliability of the simulations.

The missing mass technique was applied to the two-nucleon transfer reactions \Ni\phe\Co~and \Fe\phe\Mn~to deduce the excitation energies of $^{54}$Co and $^{50}$Mn, respectively, from the measured energy and angle of the outgoing $^3$He . 
The excitation energy spectra are shown in Fig.\,\ref{gamma}(a) and (c). The carbon background contamination is evaluated from the $^{12}$C target run with $^{56}$Ni beam. For the $^{52}$Fe beam, the same shape was considered and its amplitude was fitted to the data in the negative excitation energy region where only background contaminations contribute. Background is indicated as a thin dotted line in the excitation energy spectra.
In both cases, a broad peak centered around 0 is clearly seen together with another peak at higher energy, around 3 MeV. Simulations of the  \Ni\phe~and \Fe\phe~reactions give an excitation energy resolution of $\sigma$=600 keV and $\sigma$=650 keV respectively.  As the $^{52}$Fe beam was run after that of $^{56}$Ni, the CATS resolution was degraded due the high beam intensity already handled, resulting in a slightly degraded excitation energy resolution for $^{52}$Fe beam.    

With such an energy resolution, the contributions of the low-lying isomeric states of $^{54}$Co (197 keV, 7$^+$) and of $^{50}$Mn (225 keV, 5$^+$) cannot be disentangled from the ground-state contribution. However, the angular momentum (L) matching of the reaction favors states with L$\leq$4. Given the large L-transfer (L$\geq$5) involved for the isomeric states, they should be very weakly populated.

To check that the first peak of the \Co\, excitation energy spectrum corresponds to its ground state and thus, to a L=0 transfer, its angular distribution has been extracted. The differential cross-section is obtained from a Gaussian fit of the excitation energy spectrum between -2 and 0.6 MeV (corresponding to +1$\sigma$) for each angular bin. The fit assumes that only the ground state contributes with a fixed width determined from simulations, although  the gate includes about 30\% of the first excited state (J=1$^+$ at 937 keV) and 8\% of the J=2$^+$ state at 1445 keV.

\begin{figure}[h!]
\begin{center}
{\includegraphics[height=.20\textheight]{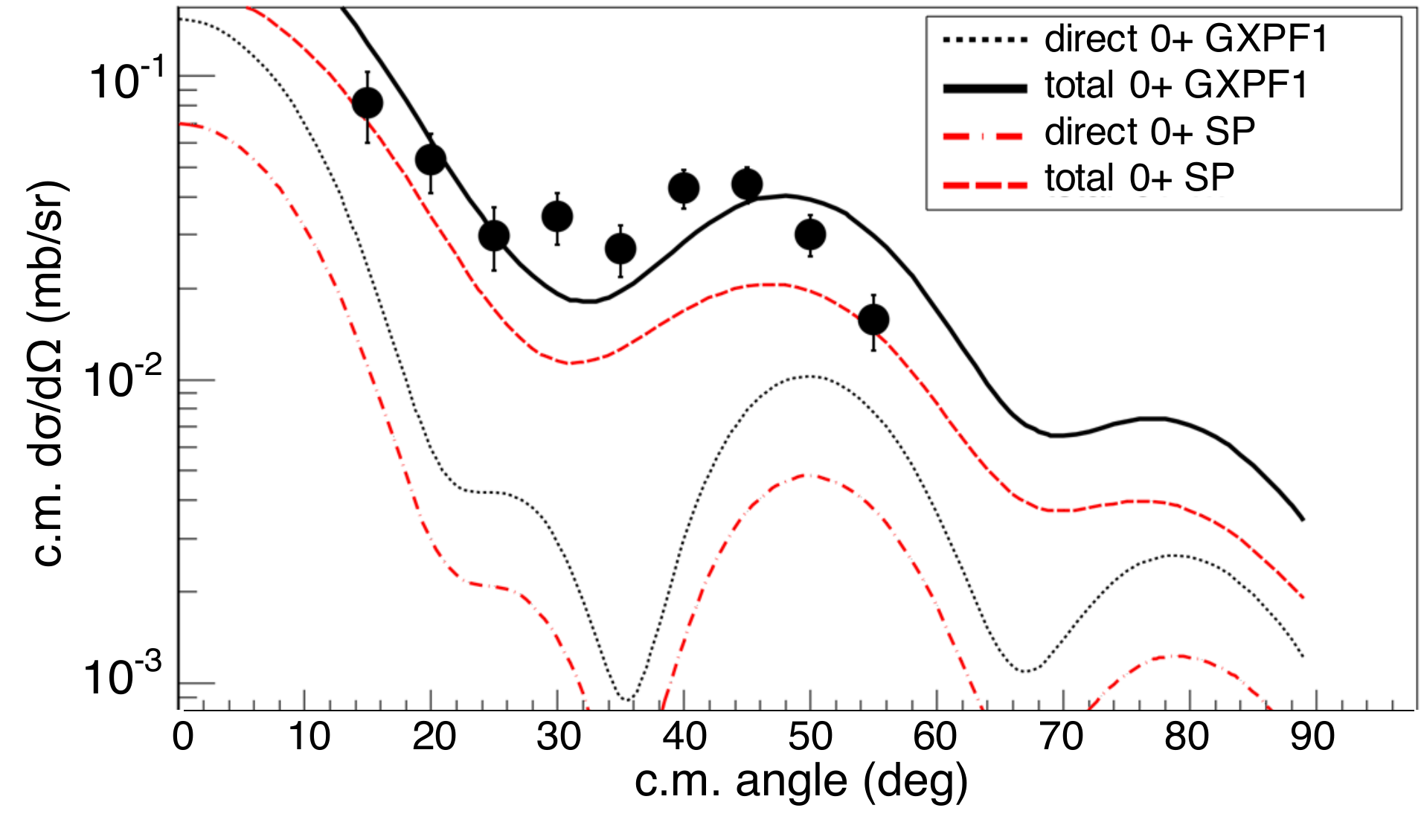}}
\caption{ (Color online) Angular distribution for \Ni\phe\Co ground state obtained in this experiment (full dots) compared with second-order DWBA calculations with GXPF1 in black (dotted line for direct transfer and full line for direct+sequential transfer) and with SP configuration in red. The error bars correspond to the statistical ones.\label{section_eff}}
\end{center}
\end{figure}

The angular distribution (see  Fig.\,\ref{section_eff}) is compared with calculations using the FRESCO code \cite{fresco}. The two-nucleon pick-up reaction (p,$^3$He) was analyzed by a finite range second-order DWBA calculation in which both sequential and simultaneous transfer are taken into account. In the simultaneous two-nucleon transfer, the prior and the post form give the same results while in the sequential (p,$^3$He) transfer, the prior and then the post form are used successively in order to eliminate the non-orthogonal terms.
The optical potentials were taken from the front-end FR2IN code \cite{fr2in}  using the Chapel-Hill 89 \cite{chap} for the proton partition while the Bechetti-Greenlees parametrization \cite{BG} was used for the $^{3}$He partition. 
The overlap functions for the heavy nuclei are described by single-particle states in a Woods-Saxon potential with the depth fitted to reproduce a state with the given binding energy. For the direct transfer contribution, the two-particle form factor is built from the two one-particle form factors and the excitation energy is shared equally between the neutron and proton. In the sequential two-nucleon transfer, only the path \Ni(p,d)$^{55}$Ni(d,$^{3}$He)$^{54}$Co is considered due to favorable binding energies in the intermediate $^{55}$Ni nucleus. The intermediate state is defined at half the binding energy difference between the initial and final states. The available energy is shared between the picked-up nucleons.
The DWBA calculations were performed for two assumptions for the $^{54}$Co ground state configuration. First, we assumed a pure single particle $\nu$\textit{f}$_{7/2}$ x $\pi$\textit{f}$_{7/2}$ configuration with no pairing (SP). Second, the two-nucleon amplitudes  (TNA) were taken from shell model calculation with the GXPF1 interaction in the \textit{fp}-shell valence space and assuming a $^{40}$Ca inert core \cite{SagToh}.  For the isovector channel, only TNA with neutron and proton in the same orbits contribute whereas in the isoscalar, spin-orbit partner cross terms also contribute.

Fig.\,\ref{section_eff} shows the DWBA angular differential cross sections for \Ni(p,$^3$He) in both cases without any normalization factor. The theoretical simultaneous and  total cross-sections for both assumptions (SP  and GXPF1) are shown. The major contribution to the cross-section for both cases comes from the sequential transfer and not from the direct transfer of a pair. A complete study of two-neutron pair transfer has shown that transfer, whether it is direct or sequential, is a quantitative probe of pairing correlations \cite{Pot13}. 
The shape of both calculations is very similar to the experimental data and confirms that the main contribution to the first peak of the excitation energy spectrum comes from an angular momentum transfer L=0, as expected for the ground-state to ground-state transfer. The overall agreement of the GXPF1 total cross-section with the experimental data is better than with the SP total cross-section, suggesting that pairing plays a significant role in the ground-state to ground-state transfer.

As the main goal of the experiment is to measure the cross-section not only to the ground-state (J=0$^+$, T=1) but also to the first excited (J=1$^+$,T=0) state (located at 937 and 651 keV for $^{56}$Ni and $^{52}$Fe respectively) and to deduce their ratio, the $^{3}$He-reconstructed excitation energy spectra alone are not sufficient.
To improve the resolving power by almost a factor of 20, the detection set-up was supplemented by four Germanium clovers of the EXOGAM array \cite{EXOGAM} placed at 55 mm around the target with front faces angles spanning from 45 to 135 degrees. The add-back photopeak efficiency of the array was measured with radioactive sources to be 8.6(1)\% at 1 MeV. After Doppler correction, the energy resolution (FWHM) for gamma-rays emitted during the decay of the populated excited states in the heavy residues is 80 keV at 1 MeV according to simulations. Particle-gamma coincidence measurements give unambiguous identification of the states populated provided that the level schemes of the residual nuclei are available. 

For \Co~and $^{50}$Mn, the low-lying excited states are well-known \cite{ENSDF}. 
The gamma spectra shown in Fig.\,\ref{gamma}(b) and (d) are gated by a 4$\sigma$-wide gate on the excitation energy centered on the J=1$^{+}$,T=0 state of interest (thus including 95\% of the contribution of this state). The corresponding gate is reported on each spectrum.
The gamma-ray lines observed are shown by vertical arrows on the gamma spectra with the same line style as in the level scheme. Within the applied gate, only the 1$^+$, 2$^+$ and 3$^{(+)}$ states are observed, as expected from the L-matching condition of the transfer reaction.  
The contributions between 100 and 300 keV correspond to the random coincidences with the gamma-rays from the decay of the implanted beams of \Ni~and \Fe. 

The gated gamma spectra of Fig.\,\ref{gamma}(b) and (d) are used to determine the cross-section for the direct feeding by the transfer reaction of the low-lying excited states in $^{54}$Co and $^{50}$Mn.
In order to extract the number of counts for each gamma-ray line, 
the spectra were efficiency corrected. An L-dependent Lorentz boost efficiency was further applied. Then, each line was fitted by a gaussian combined with a linear background.
The number of counts expected in the excitation energy spectrum for each state within the gate (reported in blue on the level schemes of Fig.\,\ref{gamma}) was deduced after subtracting the feeding from higher lying states.  

In the case of $^{54}$Co, the 3$^+$ state feeding by the transfer reaction (within the excitation energy gate) is given by its main line at 375 keV. The total population of the 2$^{+}$ state is deduced from its main line at 508 keV, relying on the tabulated gamma intensities. The deduced number of 1445 keV transitions is consistent with the measured one. Then, its direct feeding by the transfer reaction is deduced by subtracting the contribution of the 375 keV transition and is shown in blue in the level scheme. The same procedure is applied for the 1$^+$ state where the feeding from the 508 keV transition is subtracted. 

For $^{50}$Mn, the total population of the 2$^+$ state is inferred from its line at 800 keV.  The feeding from the (3) state at 2477 keV is subtracted to deduce its direct feeding by the transfer reaction. The same method is applied for  the 1$^+$ state where the feeding from the 150 keV transition is accounted for. 
The error bars on the direct feeding of the excited states by the transfer reaction are shown in parenthesis. They take into account only the statistical error. 
For both measurement, the top feeding contribution to the 1$^+$ low-lying state is important and contributes to the large error bars. As a result, the transfer cross-section to the 1$^+$ state of $^{50}$Mn could be compatible with zero.

%

With the information on the direct feeding of the excited levels, a multi-gaussian fit of the excitation energy spectra can be performed to deduce precisely the population of the ground state. 
The fit is applied between -2 MeV (to include fully the ground state) and the energy of the 1$^+$ state +2$\sigma$ (shown by the vertical red lines in Fig.\,\ref{gamma}a) and c)). The positions of the excited states are taken from the database but a global shift of the excitation energy is allowed. The width of the state is fixed as deduced from simulations. The amplitudes of the gaussians for the excited states are determined by the direct population of the states deduced from the gamma spectra and are allowed to vary within the error bars. The amplitudes are corrected from the percentage of the gaussian that is included in the fit range. The total fit is shown as a full red line in Fig.\,\ref{gamma} with each individual contribution shown with a line code corresponding to the one in the level scheme. The p-value of both fits is higher than 0.8, showing that the fit, with a limited number of parameters, reproduces well the data. The deduced ground state populations are shown in red in the level schemes together with the fit error bar.
Given the limited excitation energy resolution, the tail of the peak located between 2 and 3 MeV contributes below 2 MeV.  In the case of $^{52}$Fe\phe, the contribution of the (3) state at 2477 keV is included in the fit as it is observed in the $\gamma$-ray spectrum. Its effect is included in the fit error bars. For $^{56}$Ni, no $\gamma$-ray from this peak is identified within our gate. If we extend the gate up to 4 MeV, we observe a transition at 1650 keV that could correspond to the 3094 keV level. The contribution of this state is shown on Fig.\,\ref{gamma}(a) with a thin line. The possible overestimation of the ground-state population due to the tail of this peak is lower than 2\%.

\begin{table}[h!]
\begin{center}
\vspace{-0.cm}
\begin{tabular}{ccccc}
\hline
 &\textbf{$\sigma$(0+,T=1)} ($\micro$b)   & \textbf{$\sigma$(1+,T=0)} ($\micro$b) &  \textbf{Ratio}  \\
\hline 
\multicolumn{4}{c}{\textbf{$^{56}$Ni(p,$^3$He)$^{54}$Co}} \\
this work & 109 $\stackrel[]{stat}{\pm}$5 $\stackrel[]{sys}\pm$10 & 17 $\stackrel[]{stat}{\pm}$7 $\stackrel[]{sys}{\pm}$2 & 6.3$^{+3.1}_{-2.1}$\\
 SP & 73  & 19 & 3.8 \\
GXPF1 & 136  & 21 & 6.4\\
\hline
\multicolumn{4}{c}{\textbf{$^{52}$Fe(p,$^3$He)$^{50}$Mn }} \\
this work  & 156 $\stackrel[]{stat}{\pm}$12 $\stackrel[]{sys}{\pm}$15  & 16$^{+29}_{-16} \stackrel[]{sys}{\pm}$2 & 9.8$^{+\infty}_{-4.0}$\\
SP & 69 & 16 & 4.3\\
GXPF1 & 257 & 17  & 15.1\\
\hline
\end{tabular}
\caption{Theoretical (based on second-order DWBA calculations) and experimental cross-sections for $^{56}$Ni(p,$^3$He) and $^{52}$Fe(p,$^3$He). For cross-sections, the first error bar given corresponds to the statistical one and the second one to the systematics errors. For the ratios, the error bar is only the statistical one (see text for details).}
\label{Tab_results}
\end{center}
\vspace{-0.cm}
\end{table}

The experimental cross-sections obtained for \Ni\phe\,and \Fe\phe\, transfer to the lower lying (J=1$^+$,T=0) and (J=0$+$,T=1) states are reported in Tab.\ref{Tab_results} with their associated statistical and systematic error bars (which is dominated by the uncertainty on the target thickness). The associated theoretical predictions relying on second-order DWBA calculations using either the SP approximation ("no pairing") or the TNA from the Shell Model calculations with the GXPF1 interaction ("pairing") are also shown. 
Both for \Co and $^{50}$Mn, the contribution of $\Delta$T=1 transfer to the ground-state (J=0$^{+}$,T=1) largely dominates over the contribution of $\Delta$T=0 transfer to the excited state (J=1$^{+}$,T=0). This is reproduced by the theoretical calculations. For the deuteron-like transfer, the SP or GXPF1 calculations give similar results, showing no enhancement related to isoscalar pairing.

The weak population of the (J=1$^+$,T=0) state has also been noticed in the measurement of $^{44}$Ti($^3$He,p)$^{46}$V \cite{Fra15} but not in the two-nucleon transfer measurements performed in the \textit{sd}-shell. This observed lower cross-sections in the \textit{fp}-shell may be explained by the fact that isoscalar np pairing is more hindered by the spin-orbit effect than in the \textit{sd}-shell \cite{Ber10, Poves98, simone, Sag13}. Indeed, the np pair configurations for the isoscalar channel are not only built with the same orbitals (as for the isovector channel) but also with the spin-orbit partner orbitals. In this case, the \textit{f}$_{7/2}$ and \textit{f}$_{5/2}$ orbitals have large spin-orbit splitting which disfavors isoscalar deuteron-like pairing. As discussed in \cite{Sag13}, it reflects in the ground states of the odd-odd \textit{fp}-shell nuclei that have the configuration (J=0$^+$,T=1) except for $^{58}$Cu which has (J=1$^+$,T=0) configuration because the odd neutron and the odd proton occupy the 2\textit{p}-orbit which has smaller spin-orbit splitting. 

As for the $\Delta$T=1 transfer to the ground state, its cross-section is almost an order of magnitude larger than the deuteron-like transfer. As expected, the transfer cross-section increases at the middle of the shell where the number of contributing pairs is larger. However the measured cross-section for the ground-state to ground-state transfer \Fe\phe\Mn is lower than the predicted one. It might reflect the fact that T=1 pairing is affected by deformation \cite{Gam15}.

\begin{figure}[h!]
\begin{center}
\includegraphics[height=.23\textheight]{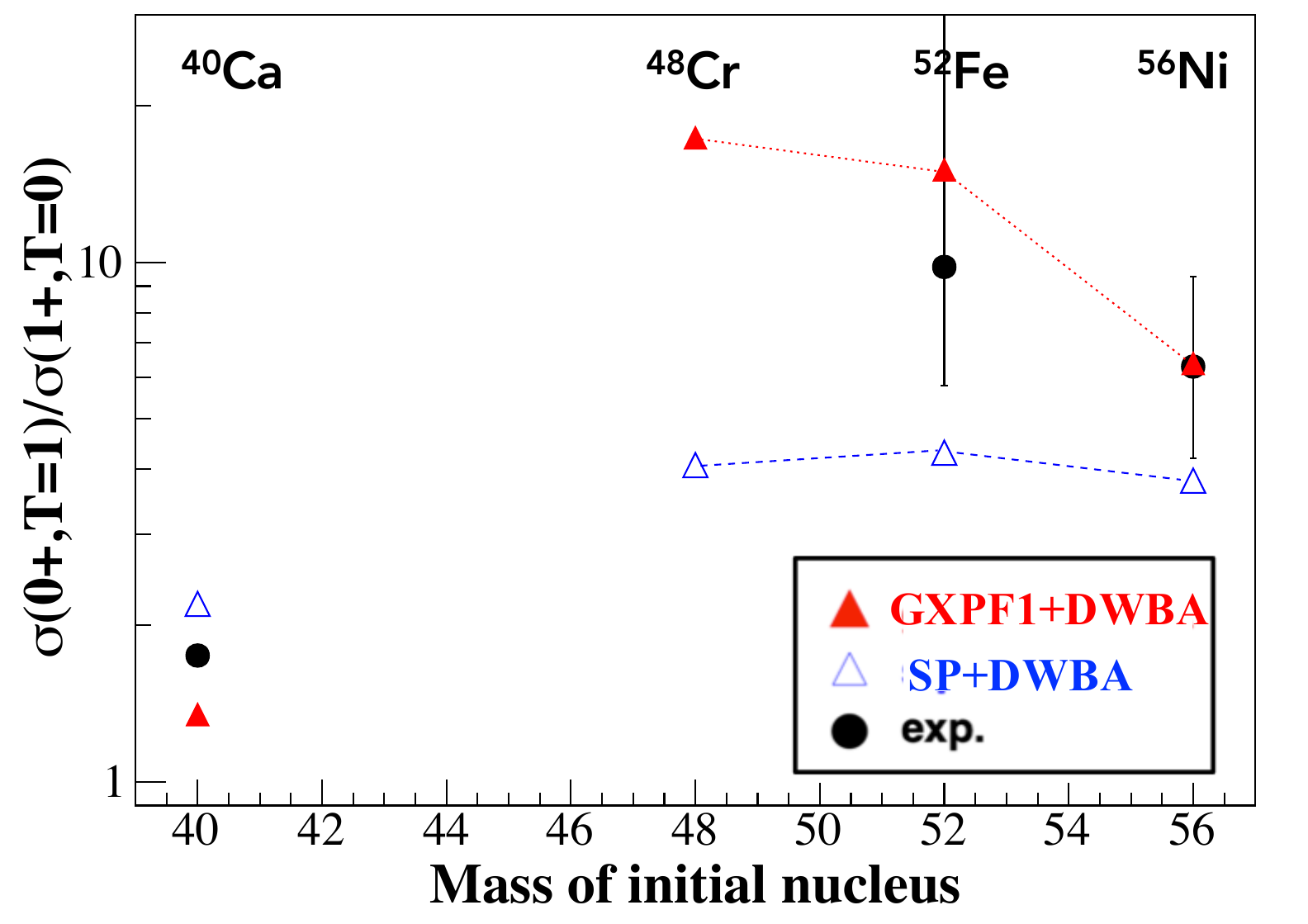} 
\caption{ (Color online) Ratio $\sigma$(0$^{+},T=1$)/$\sigma$(1$^{+},T=0$) obtained in this experiment (black dots) and for second order DWBA calculations with GXPF1(red triangle) and SP picture (blue squares). $^{40}$Ca results are taken from ref.\cite{Ayy17}.\label{results}}
\end{center}
\end{figure}

Ratios of cross-sections give valuable information on the relative strength of the isoscalar and isovector pairing. 
Given the large statistical errors on the cross-section for the (J=1$^+$,T=0) state, the usual error propagation formula based on the gaussian approximation for small errors are not valid. Following Ref.\,\cite{Sie87} for large errors, we have used the log normal distribution which results into asymmetric error bars.
The ratios of cross-sections with their error bars are reported in Tab.\,\ref{Tab_results} and on Fig.\,\ref{results} where they are compared with theoretical predictions.
For $^{52}$Fe\phe, the cross-section to the 1$^+$ state of \Mn is compatible with zero, our sensitivity being 9 $\micro$b (corresponding to one count in the gamma-ray line of this state). Thus the associated cross-section ratio $\sigma$(0$^+$,T=1)/$\sigma$(1$^+$,T=0) gives only a lower limit.  

Our measurement extends the experimental measurements reviewed in ref.\cite{Fra15} towards the \textit{fp}-shell.  
The experimental ratios are compared to the ratios deduced from second-order DWBA calculations performed with GXPF1 TNA for \Ni$~$ and \Fe$~$ as well as $^{48}$Cr\phe\,in Fig.\,\ref{results}. As a reference the ratios for the SP case are also reported. The trend of the experimental points is well reproduced by the DWBA calculations with GXPF1. Even with the large error bars, the SP, which gives an almost flat behavior, can be dismissed. When the experimental ratio is higher (lower) than the SP ratio, isovector pairing (isoscalar) dominates over the isoscalar (isovector) pairing. Both for \Co\, and $^{50}$Mn, the observed ratios point to a dominating isovector pairing. 

On Fig.\,\ref{results}, the ratio for $^{40}$Ca\phe\, measured in Ref.\,\cite{Ayy17} is also reported to complete the \textit{f}-shell and to give a reference of closed shell nucleus for comparison with the $^{56}$Ni\phe\, case. For the former, the measured ratio is 1.75, i.e. the cross-section to the (J=0$^+$, T=1) ground-state is less than twice the cross-section to the first (J=1$^+$,T=0) excited state. The experimental ratio is below the SP ratio so we can infer that, in this case, isoscalar pairing is dominant over isovector pairing. This is the opposite for $^{56}$Ni\phe\,, where the experimental ratio lies higher than the SP limit, indicating that isovector pairing dominates over isocalar pairing.

In summary, the \phe~two-nucleon transfer reactions on \Fe~(open-shell) and on \Ni\,(closed-shell) from the \textit{fp}-shell clearly show the important role of pairing correlations and, at the same time, do not provide evidence for an isoscalar deuteron-like (J=1) pairing condensate.
Either the T=0 strength is mainly concentrated in the aligned (J=J$_{max}$) configuration or, more likely, the reduced effective degeneracy to form J=1 pairs due to the spin-orbit splitting fragments the strength to higher energies with individual components not strong enough to be observed above background. 
Similar experiments on heavier N=Z nuclei in the \textit{g}-shell appear as the next logical step for further studies of the competition between isovector vs. isoscalar pairing but this will have to wait for the new generation of radioactive beam facilities.

\section*{Acknowledgments}
The authors are very grateful to GANIL staff, particularly Vincent Morel for assistance with the set-up.
The research leading to these results has received funding  from  the  European  Union under Seventh  Framework  Programme FP7 Infrastructures project ENSAR, grant agreement No. 262010.
This material is based upon work supported by the U.S. Department of Energy, Office of Science, Office of Nuclear Physics under Contract No. DE-AC02-05CH11231 (LBNL).
A.O. Macchiavelli thanks the Université Paris-Sud for support as an invited professor.

\end{document}